\documentstyle[12pt]{article}

\begin{document}
\baselineskip 24pt
\rightskip 2mm 
\titlepage
\title{Einstein-Pauli-Yukawa Paradox \\
--- What is the Physical Reality?}
\author{Guang-jiong Ni \footnotemark \\
\footnotesize Depratment of Physics, Fudan University,
Shanghai, 200433, P.R. China}
\footnotetext{E-mail: gjni@fudan.ac.cn}
\date{}

\maketitle
\centerline{December 26, 1997}
\begin{abstract}
\baselineskip 24pt What is the momentum spectrum of a particle moving in an
infinite deep square well? Einstein, Pauli and Yukawa had adopted different
point of view than that in usual text books. The theoretical and
experimental implication of this problem is discussed.
\end{abstract}

\newpage

\section{Introduction}

Looking at the title, our reader may wonder if there is some confusion
with the well known Einstein-Podolsky-Rosen (EPR) paradox? Our answer turns
out to be `no' first but `yes' afterwards.

The problem begins from an elementary example in almost every textbook on
quantum mechanics (QM). Consider a particle with mass $m$ confined in an
infinite deep square well in one dimensional space. The potential reads

\begin{equation}
V(x) = \left \{ 
\begin{array}{ll}
0, & -a<x<a \\ 
\infty, & |x| \ge a
\end{array}
\right.
\label{eq:squarepotential}
\end{equation}
The energy eigenvalue is well known: 
\begin{equation}
E_n = n^2 \frac{\pi^2 \hbar^2}{8 m a^2}, ~~~~~~(n=1,2, \cdots)
\end{equation}
while the stationary wave function of ground state $(n=1)$ reads 
\begin{equation}
\psi_{1}(x)=\sqrt{\frac{1}{a}} \cos \frac{\pi x}{2 a}  \label{eq:psi1}
\end{equation}

If one wishes to discuss the momentum representation of this state, the
usual calculation runs as follows: 
\begin{equation}
\phi_1(p)=\frac{1}{\sqrt{2 \pi \hbar}} \int_{-\infty}^{\infty} \psi_1(x)
e^{-i p x/ \hbar} {\rm d} x
\label{eq:phi1}
\end{equation}
Then the probability of momentum distribution for ground state is 
\begin{equation}
P_1(p)=|\phi_1(p)|^2= \frac{\pi \hbar^3}{2 a^3} \cos^2 (\frac{pa}{\hbar})
[p^2-(\frac{\pi \hbar}{2a})^2]^{-2}  \label{eq:P}
\end{equation}
as shown by Landau and Lifshitz \cite{Landau}.

However, there is another point of view in the literature as stressed by Tao
in a short paper \cite{Tao}. Einstein \cite{Jammer}, Pauli \cite{Pauli} and
Yukawa \cite{Yukawa} (EPY) claimed independently that, in the ground state
(\ref{eq:psi1}), there are only two sharp values of momentum ($p^{\prime}$)
each with probability $1/2$: 
\begin{equation}
P^{\prime}_1(p^{\prime}=\frac{\pi \hbar}{2a}) = P^{\prime}_1(p^{\prime}=-%
\frac{\pi \hbar}{2a}) = \frac{1}{2}
\label{eq:P'}
\end{equation}
Evidently, eq. (\ref{eq:P'}) is quite different from its continuous
counterpart, eq. (\ref{eq:P}). The question arises: which one , (\ref{eq:P})
or (\ref{eq:P'}), is correct?

In 1972, Markley claimed that (\ref{eq:P}) is correct while (\ref{eq:P'}) is
incorrect \cite{Markley}. In Ref. \cite{Tao}, Tao claimed that these two results are
both correct and incorrect and the contradiction between them actually
reflects the logical inconsistency within the framework of traditional
quantum theory. We will show that,

(i) Both calculations of (\ref{eq:P}) and (\ref{eq:P'}) are correct, but
they are discussing different things.

(ii) The contradiction or paradox between them is important, but not showing
the inconsistency of QM. Rather, it does exhibit the subtlety of a quantum
state and the correct interpretation of QM.

(iii) The above EPY paradox bears some resemblance to the famous EPR paradox
and may have some relevance to the parton model and spin crisis in high
energy physics.

\section{Two Kinds of Momentum}

The diagram of eq. (\ref{eq:P}) reminds us of the diffraction pattern by
single slit in optics with $I(x)$ versus $x$ as the substitute of $P_1(p)$
versus $p$. In an ideal experiment, if the particle confined in the
potential well (\ref{eq:squarepotential}) can be liberated along the $z$
direction into a broad screen along $x$ direction, we would see the number
distribution probability of particle ($I(x)$ vs $x$) just like eq. (\ref
{eq:P}). This means that the $p$ in eqs. (\ref{eq:phi1}) and (\ref{eq:P}) is
the momentum defined in infinite $x$ space, the relation between $\psi _1(x)$
and $\phi _1(p)$ is a Fourier transform. Being the eigenvalue of operator 
\begin{equation}
\hat{p}=-i\hbar \frac \partial {\partial x},~~~~~~(-\infty <x<\infty )
\label{eq:hatp}
\end{equation}
$p$ has a continuous spectrum.

On the other hand, eq. (\ref{eq:P'}) is a result coming from the direct
observation of eq. (\ref{eq:psi1}) as 
\[
\psi_1(x) = \frac{1}{2} \sqrt{\frac{1}{a}} \{ \exp (i \frac{\pi x%
}{2 a})  +\exp (-i \frac{\pi x}{2 a}) \} 
\]
with operator 
\begin{equation}
\hat{p}^{\prime} = -i \hbar \frac{\partial}{\partial x}, ~~~~~~~~~(-a<x<a)
\label{eq:hatp'}
\end{equation}
Hence, $\hat{p}^{\prime}$ has a discrete spectrum with periodical boundary
condition in $4a$ interval,
\begin{equation}
p^{\prime}_n =\frac{n \pi \hbar}{2 a}, ~~~~~~~~(n= \pm 1, \  \pm 2, \cdots)
\label{eq:p'eigen}
\end{equation}
and the relation between $\psi_1(x)$ and $\phi_1^{^{\prime}}(p^{%
\prime})$ ($P_1^{\prime}(p^{\prime})=|\phi_1(p^{\prime})|^2$) is a Fourier
series.

The difference between $\hat{p}$ and $\hat{p}^{\prime}$ is shown at eqs. (%
\ref{eq:hatp}) and (\ref{eq:hatp'}) by the definition range of $x$. It can
be displayed even more clearly if they are expressed in other
representation, say, $\hat{p}^{\prime}$ can be expressed by an infinite
matrix with matrix element (\ref{eq:p'eigen}) whereas $\hat{p}$ cannot.
Furthermore, there is a momentum conservation law for $p$ and
correspondingly an uncertainty relation reads 
\begin{equation}
\Delta x\Delta p\ge \frac \hbar 2  \label{eq:uncertainty}
\end{equation}
whereas the momentum conservation law related to $p^{\prime }$ is a discrete
one with possible momentum exchange with the lattice (Umkrapp process in
crystal) and there is no uncertainty relation like eq. (\ref{eq:uncertainty}%
). Only at large quantum number ($n\gg 1$) case,
$\phi _n^{\prime}(p_n^{\prime })$ approaches to $\phi (p)$.

As mentioned above, the spectrum of $p$ along $x$ direction could be
measured if the particle is deconfined via the $z$ direction. On the other
hand, one can imagine the measurement of $p^{\prime }$ spectrum `locally' if
the observer and apparatus are also confined in the deep potential well.

So in our point of view, both calculations of (\ref{eq:P}) and (\ref{eq:P'})
are correct because they are talking about two different kinds of momentum.

\section{What is the real momentum of particle?}

Let us insist on querying that `which one, $p$ or $p^{\prime }$, is the real
momentum of particle before the measurement is performed'? Then we are facing
a dilemma. Logically, eqs. (\ref{eq:P}) and (\ref{eq:P'}) can not all be
correct since they are different.

It seems to us that the only way out of the dilemma is not taking the
question seriously. Because if taking seriously, neither $p$ nor $p^{\prime}$
is really existing before the measurement is made.

In Dirac notation, a quantum state, {\it e.g.}, the ground state under
consideration, is denoted by an abstract state vector $|\psi_1>$. In
Heisenberg picture, there is no description either $x$ or $t$ in $|\psi_1>$.
Only after some representation is chosen, can it get some description. For
instance, if we choose the eigenvector of the position $|x, t>$ as the base
vector and take the contraction (projection) of $|\psi_1>$ with $|x,t>$,
we obtain the wave function in configuration space: 
\begin{equation}
\psi_1(x,t) = <x,t|\psi_1>
\label{eq:xwave1}
\end{equation}
as shown in eq. (\ref{eq:psi1}) with a factor $\exp (-i E_1 t/\hbar)$
omitted.

Alternatively, we can choose the eigenvector of momentum $p$, $|p,t>$, as the
base vector to get the wave function in momentum ($p$) representation: 
\begin{equation}
\phi_1(p,t) = <p,t|\psi_1>
\label{eq:pwave1}
\end{equation}
or in $p^{\prime}$ representation: 
\begin{equation}
\phi_1^{^{\prime}}(p^{\prime}, t) = <p^{\prime}, t|\psi_1>
\label{eq:pwave1'}
\end{equation}

The three kinds of wave function, (\ref{eq:xwave1}) --- (\ref{eq:pwave1'}),
are three different descriptions for a same quantum state $|\psi_1>$. No one
among the three is more fundamental than others.

The wave functions in QM are not observable. But they are very useful in linking
the even more abstract state vector, say $|\psi_1>$, to the potential possible
outcome in experiments if the latter are really performed on the state.
The variable, $x$ in (\ref{eq:xwave1}), or $p$ in (\ref{eq:pwave1}),
or $p^{\prime}$ in (\ref{eq:pwave1'}) is just to characterize what kind of experiment,
for position ($x$) measurement, or for $p$ ($p^{\prime}$) measurement, etc. we are
going to use. Therefore, $p$ (or $p^{\prime}$) in wave function is by no means the
momentum of particle
before the measurement. Similarly, the variable $x$ is also not the position of
`point particle' before the measurement. It is just because they are the
outcome of `fictituous measurement', so the probability amplitudes, i.e.,
the wave functions, contain the $i=\sqrt{-1}$ which is unobservable.

In short, a quantum state in abstract sense contains no information. Certain
information is gained only after certain measurement is made. However, the power
of QM lies in the fact that we can predict what will happen before certain
measurement is performed if the knowledge about the environment of particle, e.g. eq.
(\ref{eq:squarepotential}), is known.

\section{The Possible relevance with EPR Paradox and Their Experimental Implications}

In 1935, based on the belief of `physical reality', Einstein, Podolsky and Rosen
(EPR) raised their famous paradox. Since then, especially after the study on Bell
inequality, the various EPR experiments have been achieving an important conclusion
that the prediction of QM is correct whereas the existence of `local hidden
variable' is incompatible with experimental results (see refs. \cite{Tapster},
\cite{Tittel}). 
The subtle entanglement in two-particle system, to our understanding,
implies that before
the quantum coherence of entangled state is destroyed by measurement, no information
about one individual particle can be isolated from the other. Actually, such kind of
information does not really exist before the measurement is made.

Now the EPY paradox discussed in this paper, according to our understanding, has the
similar implication as EPR paradox but for one particle state.

The discussion on EPR or EPY paradox is by no means a pure academic one. We are facing
more and more quantum subtlety after the establishment of QM in 70 years ago.
The fact that the canonical momentum $\hat{p}$ is not an observable becomes
more evident when the particle with charge q is coupled to the gauge field,
say, the vector potential of electromagnetic field, $\vec{A}$. Then
$\hat{\vec{p}}$ should be replaced by the kinematic momentum
$\hat{\vec{\pi}}$,
\begin{equation}
\hat{\vec{p}} \rightarrow \hat{\vec{\pi}} = \hat{\vec{p}} -
\frac{q}{c}\vec{A}
\end{equation}
In a gauge transformation, $\hat{\vec{p}}$ is not invariant whereas
$\hat{\vec{\pi}}$ is. The latter is the quantum counterpart of classical
mechanical momentum $m \vec{v}$. However, the explanation of
$\hat{\vec{\pi}}$ is obscured by the noncommutativity relation
\begin{equation}
[\hat{\pi}_{i}, \hat{\pi}_{j}] = i \hbar \frac{q}{c} \varepsilon_{ijk} B_{k}
\end{equation}
($\vec{B}$ is the magnetic field). An important case with degeneracy is
electron ($q=-e<0$) moving in a plane with a constant $B$ along $z$ direction.
There are three kinds of wavefunctions describing the motion of electron.

1. Landau gauge (\cite{Landau}), $A_{x}=-By, A_{y}=A_{z}=0$.
\begin{equation}
\psi^{(1)}_{n}(x,y)= \exp(i p_{x} x / \hbar) \exp[-(y-y_{p})^2/2 a^2]
{\rm H}_{n}((y-y_{p})/a)
\label{eq:psi1n}
\end{equation}
\[
E_{n}=(n+1/2) \hbar \omega_c, ~~~~~(n=0,1,2,\cdots)
\]
\[
y_p=\frac{c p_x}{e B}, ~~~ a=\sqrt{\frac{\hbar}{m \omega_c}}, ~~~~
\omega_c=\frac{e B}{m c}
\]
For fixed nth Landau level, the state degeneracy $G$ is reflected in $y_p$.
\[
G=\Phi/\phi_0, ~~~~(\Phi=BL_{x}L_{y}, \phi_{0} = h c/e)
\]
$L_{x}$ and $L_{y}$ are the width of plane.

2. Symmetric gauge, $A_{x}=-\frac{1}{2} B_{y}, A_{y}=\frac{1}{2} B_{x}$
\cite{Huang}.
\begin{equation}
\psi^{(2)}_{nL}(z)=const (z-\frac{\partial}{\partial z^{*}})^{L}
(z^{*}-\frac{\partial}{\partial z})^{n} \exp(-z^{*} z)
\label{eq:psi2nL}
\end{equation}
\[
z=\frac{1}{2 a} (x+i y)
\]
while $E_{n}=(n+1/2) \hbar \omega$ as before, the quantum number
$L=0,1,2, \cdots$ shows that for fixed $n$ the wavefunctions are peaked at
concentric circles, with equal area between successive circles.
Each ring between circles supports one $\phi_{0}$, giving $G$ as before.

3. For $n=0$ case, the $L$ degeneracy can be transformed into wavefunctions
described by
\begin{equation}
\psi^{(3)}_{0 i}(z-z_{i})=const \exp(z z^{*}_{i}-z^{*} z_{i})
\exp(-|z-z_{i}|^2)
\label{eq:psi30i}
\end{equation}
with vortices center $z_{i}$ located at regular lattice (\cite{Huang}).
Each vortics carries a unit of $\phi_{0}$, giving $G$ as before.

At first sight, the microscopic pictures of electron motion in three kinds of
wavefunction are quite different. But acturally, none of them is more
fundamental than others. One can only say that the eq. (\ref{eq:psi1n}) is
most suitable for discussing the quantum Hall effect when electric voltage $V$
is applied along the $y$ direction and the current $I_{x}$ along $x$ axis
is really measured. A simple substitution
$y_{p}=\frac{c}{e B} (p_{x}-\frac{m c V}{B L_{y}})$ gives
$I_{x}=-\frac{e}{m}\int<\pi_{x}> {\rm d}y=-\frac{e c V}{\Phi}$.

For another example, a proton near its ground state is comprising of three
(valence) quarks
(uud). But when it is under high energy collision, its behavior can be better
analyzed by parton model, i.e., by assuming that it is composed of numerous sea
quarks (u$\bar{\rm u}$, d$\bar{\rm d}$, s$\bar{\rm s}$, etc.) and gluons besides
the three valence quarks. Perhaps one would say that if we can find a unified
theory even more basic than QCD, then all present models could be derived. We don't
think so. A proton is infinite in essence, it has various aspects in various
experiments. Not only dynamics, but also its ingredients are depending on the character of experiment, {\it i.e.}, on what we are looking for.

In recent years, there is a puzzle in high energy physics, called as `spin crisis'.
It seems that only a part of proton spin can be ascribed to that of valence quarks.
What is the other part of proton spin coming from? Then the analysis becomes quite
confused as how to define the angular momentum of quarks and gluons \cite{Wang}.  
We are far
from the experts in this field. However, the similarity between the confinement of
quark (gluon) in proton and that of the particle in an infinite deep square well
as in this paper also deserves consideration. When we are talking about the momentum of
a quark inside a proton, is it bearing some resemblance to $p$ in eq. (\ref{eq:P})
or $p^{\prime}$ in eq. (\ref{eq:P'})? Alternatively, either $p$ or
$p^{\prime}$ may be suitable from the beginning, but the difference in various outcome
of analysis may be stemming from different approximation which is inevitable in
analysis. Anyway, we had better become aware of the possibility that sometimes we
may talk about something too seriously.

\section{What Is the `Physical Reality'?}

We are now in a position to try to answer the question raised by EPR: `What is the
physical reality?' It seems to us that a more rigorous philosophical definition
is needed. A `thing' should be defined at two levels. An object, when it is
independent of the consciousness of mankind and before the measurement is made,
could be called as `thing in itself'. It is something absolute in nature and
containing no information. In QM, it is denoted by a quantum state $|\psi>$
separated approximately from its environment. Then after certain measurement is
performed, it is
turned into `thing for us', reflecting a series of experimental data. It is then
something relative in nature. Sometimes, we call it `phenomenon'. As Wheeler said:
`No phenomenon is phenomenon until it is an observed phenomenon'. The
wavefunction in QM is just playing the role for connecting the two levels
of matter via the fictitious measurement. Eventually,
we will be convinced by the Chinese saying `Oneness of heaven and man'.
\\
\\
\centerline{\Large \bf Acknowledgement}
{  } The author wishes to thank Profs. H. Chen, X-x Dai, C-d Gong, G-c Guo,
P. Jiang,
R-k Su, Z-y, Tao, F. Wang, B-w Xu, S-q Ying and X-y Zeng for discussions. This work
was
supported in part by the National Science Foundation of China.


\begin{thebibliography}{9}
\bibitem{Landau}  L.D. Landau and E. M. Lifshitz, {\it Quantum Mechanics,
	Non-relativistic Theory}, (3rd Ed.), Oxford : Pergamon Press, 1977, P.65.

\bibitem{Tao}  Tao Zongying, Acta Photonica Sinica, $\underline{26}$(1997) No. 9,
	769 $\sim$ 770.

\bibitem{Jammer}  Max Jammer, {\it The Philosophy of Quantum Mechanics, the
Interpretation of Quantum Mechanics in Historical Perspective}. New York:
Wiley, 1974: 218 $\sim $ 219, 132 $\sim $ 136.

\bibitem{Pauli}  Wolfgang Pauli, {\it Pauli Lecture on Physics, Vol 5: Wave
Mechanics}. Cambridge: (Mass) MIT Press, 1973, 23 $\sim $ 24.

\bibitem{Yukawa}  H. Yukawa (Ed), {\it Quantum Mechanics} (Vol I, 2nd Ed),
(in Japanese), Tokyo Yan-Bo Bookshop, 1978: 113.

\bibitem{Markley}  F.L. Markley, Amer. Jour. Phys. 40 (1972) No. 10, 1545 $%
\sim $ 1546.

\bibitem{Tapster} P. R. Tapster, J. G. Rarity and P. C. M. Owens,
Phys. Rev. Lett. {\bf 73}, 1923(1994).

\bibitem{Tittel} W. Tittel, J. Brendel, B. Gisin, T. Herzog, H. Zbinden
and N. Gisin, preprint, Internet, quant-ph/9707042, to appear in Phys.
Rev. A.

\bibitem{Huang} K. Huang, MIT preprint CTP \#1265(1985),
lecture at Fudan University, May 27, 1985.

\bibitem{Wang} X-s Chen and F. Wang, preprint, Gauge Invariance and
Hadron Structure.

\end{thebibliography}
\end{document}